\documentclass[twocolumn,showpacs,aps]{revtex4}
\usepackage{graphicx}
\usepackage{subfigure}
\usepackage{subfig}
\usepackage{amsmath}
\usepackage{amssymb}

\begin{document}
\title{The Iterative Unitary Matrix Multiply Method and Its Application to Quantum
Kicked Rotator}
\author{Tao Ma}
\affiliation{Department of Modern Physics, University of Science and
Technology of China, Hefei, PRC}
\date{\today}



\begin{abstract}
We use the iterative unitary matrix multiply method to calculate the
long time behavior of the resonant quantum kicked rotator with a
large denominator. The delocalization time is exponentially large.
The quantum wave delocalizes through degenerate states. At last we
construct a nonresonant quantum kicked rotator with delocalization.
\end{abstract}

  \maketitle

\emph{Introduction.}---The quantum kicked rotator (QKR)
\cite{Casati1979}, which describes a periodically kicked rotator, is
one of the most studied model of quantum chaos
\cite{stockmann1999qci}. The classical correspondence of QKR is the
standard map \cite{chirikov1979uim, greene1979mds}. Classically, the
energy of the rotator grows without a limit. But to a quantum
rotator, if the kick frequency and the rotator frequency is
commensurate, QKR delocalizes in the momentum space and if
incommensurate, QKR generally localizes \cite{Casati1979}. Fishman
\emph{et al} explained the classical and quantum difference by
transforming QKR into an Anderson localization problem
\cite{Fishman82}.

In the paper, we try to understand how the delocalization of
commensurate cases happens. This is important for several reasons.
First, the commensurate (incommensurate) case is described by a
rational (irrational) number. The qualitative statement that
delocalization happens to the commensurate cases is correct but
incomplete. We want to gain a quantitative understanding. Second, no
physical quantity is rational or irrational. A physical quantity has
only several significant digits, while the distinction between
rational and irrational numbers depends on infinite significant
digits. An infinitesimal error can change a rational (irrational)
number into an irrational (rational) number. While we expect the
system changes little from our experiences of studying physics as
was emphasized by Hofstadter \cite{Hofstadter76}. To recognize and
reconcile the conflict is one aim of quantum chaos. Third, Fishman
\emph{et al}'s result \cite{Fishman82} seems to tell us localization
happens to all the incommensurate cases. Is there at least one
incommensurate case for which delocalization happens? Casati
\emph{et al} has derived a quantum Lyapunov equation to describe the
difference between the dynamics of commensurate and incommensurate
cases \cite{Casati1984, TaoMa2007GeneralI}. Based on the formula,
Casati \emph{et al} claimed there are some incommensurate cases of
delocalization \cite{Casati1984}. But their argument is problematic
\cite{TaoMa2007GeneralI} from the perspective of the exponentially
large delocalization time discovered in the paper.

In the paper, we prove by numerical calculation for the commensurate
case with a large denominator, the delocalization time is
exponentially large. Such a large denominator effect is explained by
the degenerate perturbation theory, which is based on the
observation that degenerate states are the delocalization path.
Localization of incommensurate cases can be understood to be caused
by the large denominator effect. The large denominator effect and
the quantum Lyapunov equation \cite{Casati1984, TaoMa2007GeneralI}
partially reconcile the conflict between commensurate and
incommensurate cases and naturally lead to an incommensurate case of
delocalization. This partially solves the problem: to find an
incommensurate case of delocalization, posed by Casati \emph{et al}
\cite{Casati1984, Casati1986} and gives a counterexample to
Fishman's argument \cite{Fishman82}, although a very weak one.

\emph{Numerical methods.}---For a system with a periodical
Hamiltonian, the unitary operator of one period is the Floquet
operator $F$. The unitary operator of $2^N$ periods is $F^{2^N}$.
\begin{eqnarray} \label{eqIUMM}
F^{2^N}&=&(F^{2^{N-1}})^2; \nonumber\\
F^{2^{N-1}}&=&(F^{2^{N-2}})^2; \nonumber\\
   &\cdots&  \nonumber\\
F^{4} &=& (F^{2})^2.
\end{eqnarray}
From Eq.~(\ref{eqIUMM}), we can calculate $F^{2^N}$ from $F$ by
iteratively multiplying the unitary matrices for $N$ times. This
method is referred as the iterative unitary matrix multiply method
(IUMM). It is impossible to calculate very long time behavior of QKR
using the usual fast Fourier transform method \cite{Fishman82}. IUMM
is actually the same method as direct diagonalization or the matrix
vector multiply method used in the original paper \cite{Casati1979}
of QKR. See the section III and IV of \cite{Casati1979}.

\emph{Calculation results.}---The Hamiltonian of QKR is
\begin{equation}
H=-\frac{1}{2}\hbar^{2}\frac{\partial^2}{\partial\theta^{2}}+k
\cos\theta\sum_{n=1}^\infty \delta(t-n\tau),
\end{equation}
where $\hbar$ is the Planck constant, $\tau$ the kick period and $k$
the kick strength. The matrix element of $F$ is
\begin{equation}
F_{nm}=\langle n| F|m \rangle= \exp(-i \hbar\tau \frac{m^2}{2})
i^{m-n} J_{n-m}(\frac{k}{\hbar}),
\end{equation}
where $|n\rangle=\frac{1}{\sqrt{2\pi}}e^{in\theta}$. We apply IUMM
to QKR. In the calculation $\hbar=1$, $k=1$ and
$\tau=\frac{2\pi}{q}=\frac{\pi}{10}$. The initial state is
$|0\rangle$. This is the commensurate/resonant case with a large
denominator $q=20$. $F_{nm}=F_{n+20, m+20}$ for every $n$ and $m$.
The rotator will delocalize in the future, nevertheless it
delocalizes very slowly.

\begin{figure*}
\begin{center}
   \begin{minipage}{17.0 cm}
   \includegraphics[width=16.0cm]{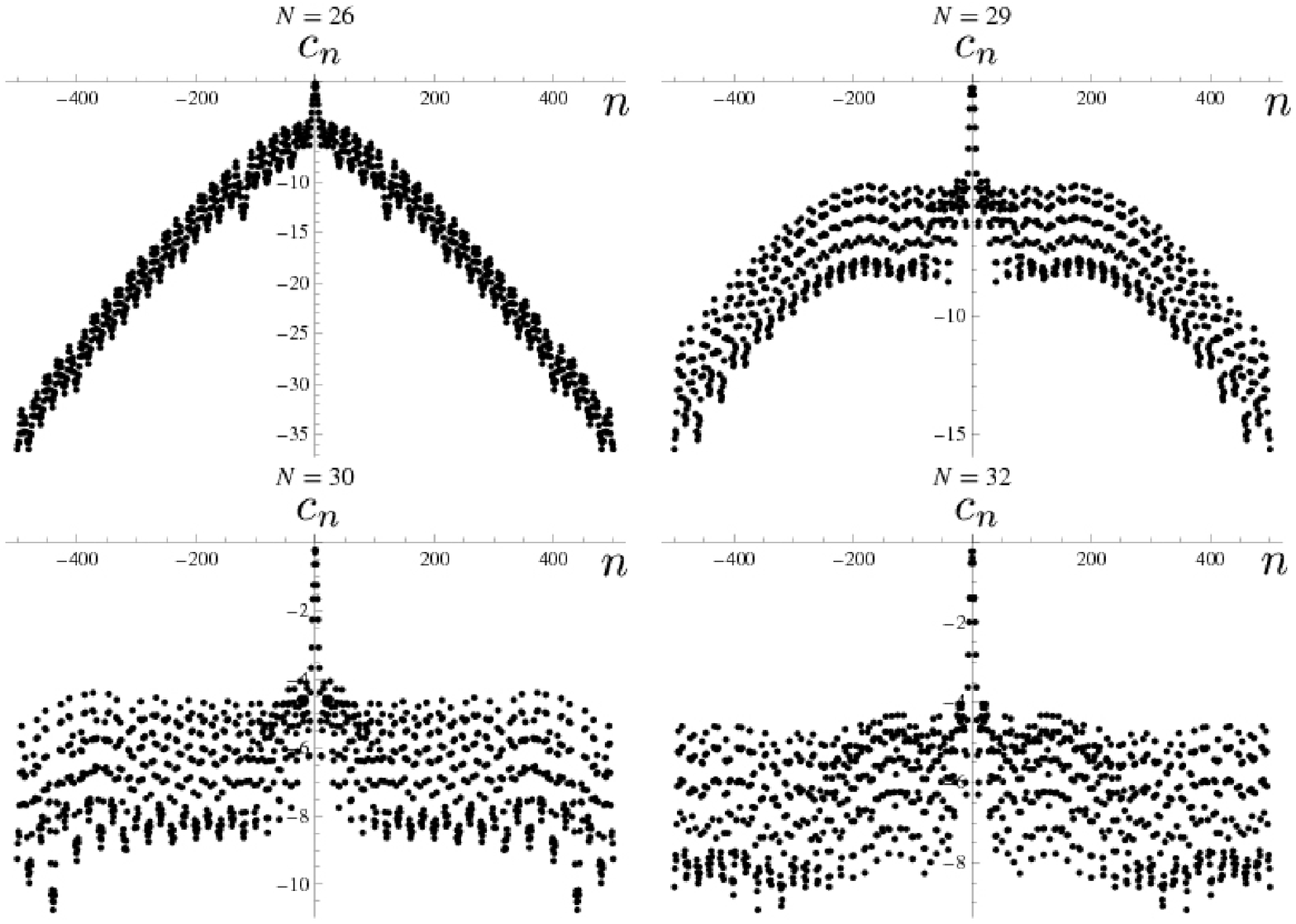}
   \includegraphics[width=16.0cm]{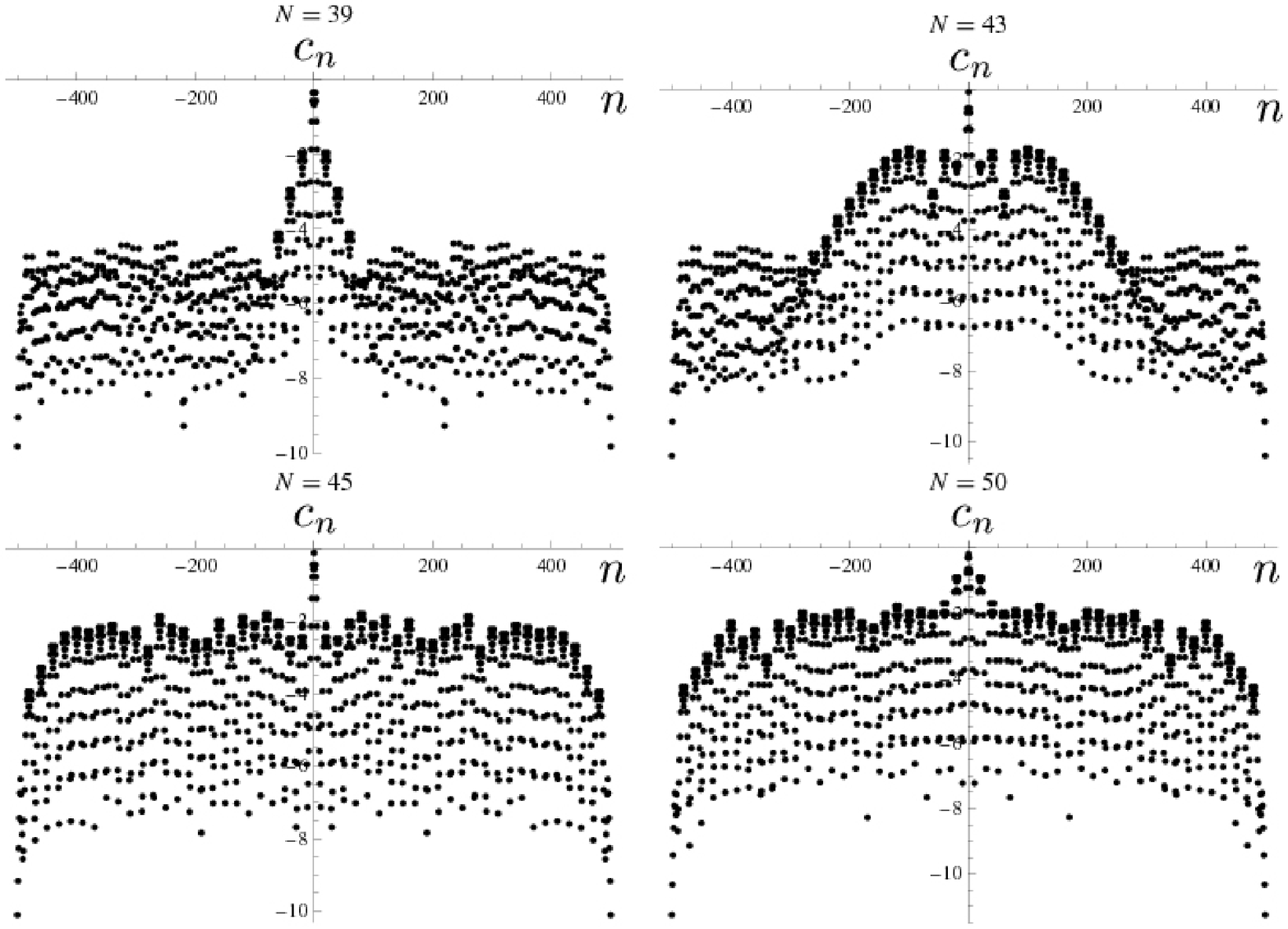}
\caption{QKR wave function at different time. $N$ is at time
$2^N\tau$. $n$ is $|n\rangle$ and $c_n$ is the base-10 logarithm of
the absolute value of the wave function on $|n\rangle$. $n$ is from
$-500$ to $500$ in our calculation.}
\end{minipage}
\end{center}
\end{figure*}

\begin{figure*}
\begin{center}
   \begin{minipage}{17.0cm}
   \includegraphics[width=16.0cm]{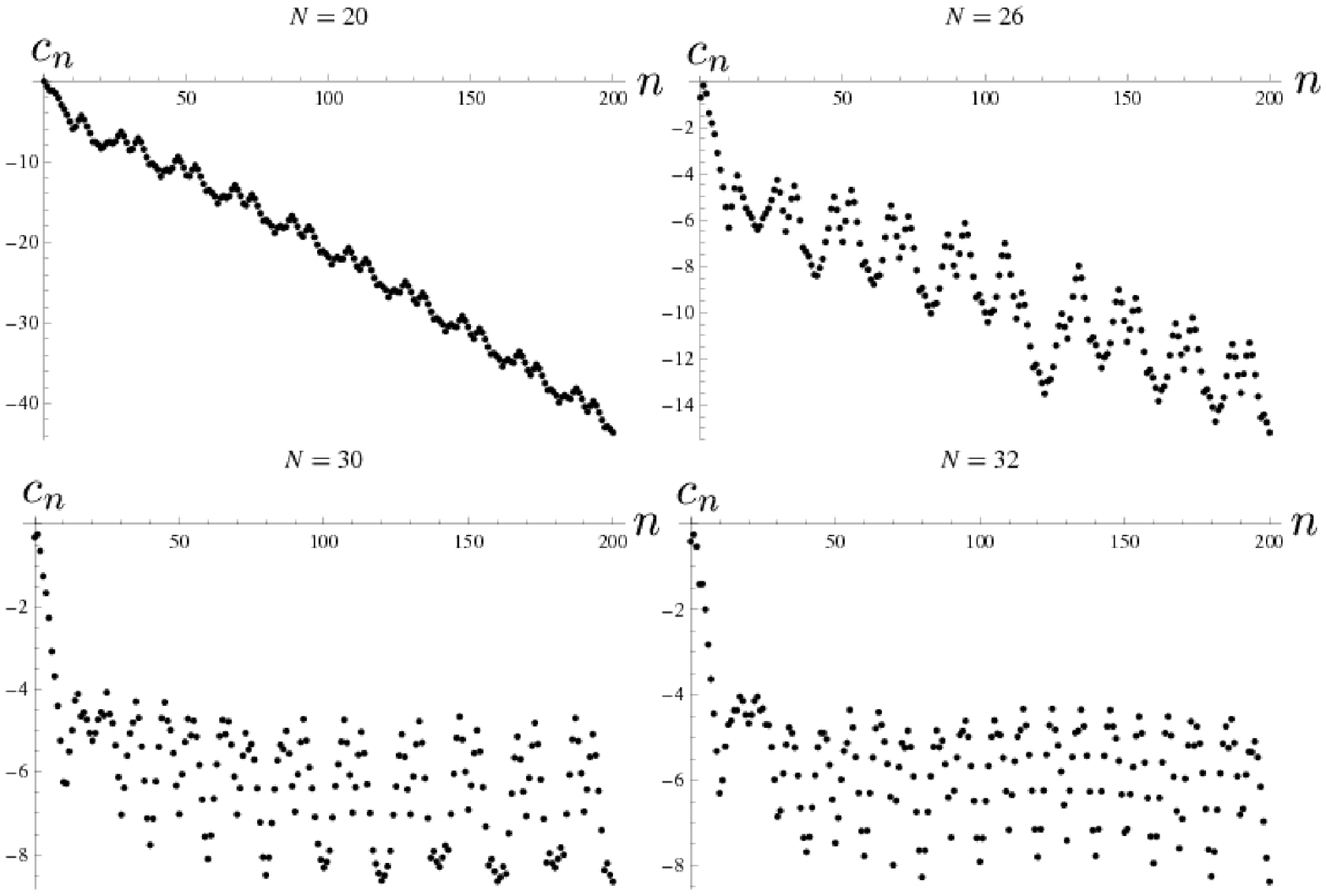}
   \includegraphics[width=16.0cm]{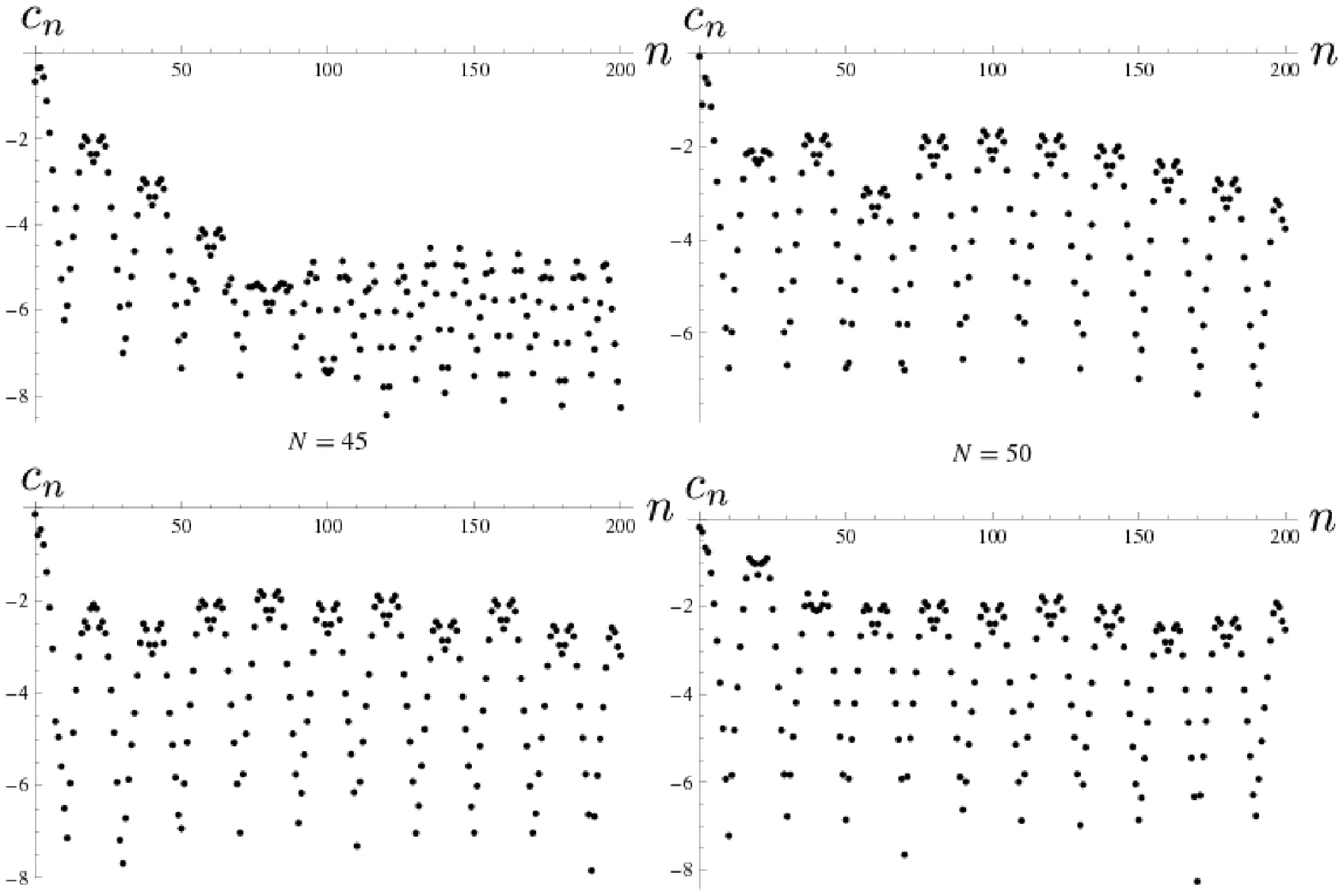}
\caption{Clearer figures of QKR wave function. Note the changes of
peaks and valleys from $N=26$ to $N=43$. The peaks indicate what $|n
\rangle$s actively contribute to the wave propagation.}
\end{minipage}
\end{center}
\end{figure*}

FIG. 1 and FIG. 2 show the same calculation results. The
distribution of QKR wave function is clearer in FIG. 2. Before
$N=20$ (the time is $2^{20} \tau=1.05\times 10^6\tau$), QKR does not
delocalize at all. To a resonant case, this is unexpected. At
$N=20$, peaks (local maxima) and valleys (local minima) appear.
Naively one expects, from $N=20$ to 50, peaks should be at
$|n=20\times I \rangle$s, where $I$ is an integer, because such
states are resonant with the $|0\rangle$. Actually $|n=20\times I
\rangle$s are always valleys. Before $N=39$, the peaks are at
$|n=10\times O \pm 3\rangle $s, where $O$ is an odd integer, such as
$n=13, 27, 33, 47, 53, \cdots$. At $N=50$, the peaks are at $|n=20
\times I \pm 3\rangle$s, such as $n=17, 23, 37, 43, \cdots$.

At $N=26$, the triangle like wave function outside $n=0$ forms and
at $N=30$ it flattens. The height of the flattened wave function is
approximately $10^{-4}$. The main distribution is still at
$|0\rangle$. From $N=30$ to 39, the needle like wave function around
$n=0$ becomes a triangle like one. From $N=39$ the triangle expands
and at $N=45$ the wave function flattens again. The height of the
QKR wave function is approximately $10^{-2}$. At $N=50$, a triangle
like wave function forms again.

Around $n=0$, the circulation: needle $\rightarrow$ triangle
$\rightarrow$ expanded triangle $\rightarrow$ flattened, drives the
whole delocalization process. Will the wave function around $n=0$ be
totally flattened by one more circulation or several more after
$N=50$? Will $| n=20 \times I \rangle$s finally become the only
peaks when $N$ is very large and how? We do not know! Our
calculation overflows around $N=60$, which may be caused by the
ununitarity of the truncated Floquet operator.

At what $N$ should the wave function be considered as delocalized?
At $N=30$, there is still lots of distribution of the wave function
at $n=0$. $N=45$ is more proper than $N=30$. We define
$T_{\tau/2\pi}\times \tau$ as the delocalization time of QKR with
the period $\tau$. For simplicity, we also refer $T_{\tau/2\pi}$ as
the delocalization time. For $\tau=2\pi/q$, QKR delocalizes
exponentially slowly. We estimate $T_{1/q} \approx \exp(cq/k)$,
where $c$ is a factor that depends weakly on $k$ and $q$. If we
assume QKR is delocalized at $N=45$, $e^{20c} \tau \approx 2^{45}
\tau$ and $c \approx 1.56$.

\emph{Delocalization path and degenerate perturbation theory.}---The
delocalization time can be estimated from the degenerate
perturbation theory. The sequence $\{ -\frac{1}{40} n^{2} \text{Mod}
(1) \}_{n=0,1,\cdots,20}$, which are the phases of $F_{nn}$ divided
by $2\pi$, is
\begin{equation}
\begin{split}
\{0,&\frac{39}{40},\frac{9}{10},\frac{31}{40},\frac{3}{5},\frac{3}{8},\frac{1}{10},\frac{31}{40},
\frac{2}{5},\frac{39}{40},\frac{1}{2}, \\
&\frac{39}{40},\frac{2}{5},\frac{31}{40},\frac{1}{10},\frac{3}{8},
\frac{3}{5},\frac{31}{40},\frac{9}{10}, \frac{39}{40},0\}.
\end{split}
\end{equation}
In a period, there is four $\frac{39}{40}$s, four $\frac{31}{40}$s,
two $\frac{9}{10}$s, two $\frac{3}{5}$s, two $\frac{2}{5}$s, two
$\frac{3}{8}$s, two $\frac{1}{10}$s, one $\frac{1}{2}$ and one $0$.
The quantum wave is easier to propagate between $\frac{31}{40}$s or
between $\frac{39}{40}$s. So $|n=20\times I\rangle$s are valleys
from $N=20$ to $50$. When $n=1, 9, 11, 19, 21, 29, 31,39$, the
phases are $\frac{39}{40}\times 2\pi$. The intervals between two
degenerate states are 1 and 8. When $n=3, 7, 13, 17, 23, 27, 33,
37$, the phases are $\frac{31}{40}\times 2\pi$. The intervals are 4
and 6. So the wave is easiest to propagate between $n=3, 7, 13, 17,
23, 27, 33, 37$, which are peaks in FIG. 2. But we do not know why
peaks are at only some of the $|n=10\times I\pm 3\rangle$s. Peaks
even change from the $|n=10\times O\pm 3\rangle$s to the
$|n=20\times I\pm 3\rangle$s as we discussed above.

From the degenerate perturbation theory, if the wave propagates
through the path $|0\rangle \rightarrow |20\rangle$, $F$ is
approximated by
\begin{equation}\label{degenerate1}
F_{\text{appr}}= \left(
\begin{array}{ll}
 F_{0,0} & F_{0,20} \\
 F_{20,0} & F_{20,20}
\end{array}
\right)
=
\left(
\begin{array}{ll}
 J_0(k) & J_{-20}(k) \\
 J_{20}(k) & J_0(k)
\end{array}
\right).
\end{equation}
The eigenvalues of $F_{\text{appr}}$ is $J_0(1) \pm J_{20}(1)$. So
after approximate $J_0(1)/J_{20} = 1.98\times 10^{24}$-time kicks,
the wave function will be transferred from $|0\rangle$ to
$|20\rangle$. This is far larger than $2^{45} = 3.52\times 10^{13}$
of our numerical result. A more exact estimate has to take into
account other degenerate states. The quantum wave can propagate
through the path $|0\rangle \rightarrow |1\rangle \rightarrow
|9\rangle \rightarrow |11\rangle \rightarrow |19\rangle \rightarrow
|20\rangle$. The states contributing to the wave propagation are
mainly these states. So $F$ is approximated by $F_{\text{appr}}$,
which only considers the states in the delocalization path.
\begin{equation}\label{degenerate2}
F_{\text{appr}}=\begin{pmatrix}
                    F_{0,0} & F_{0,1} &   &   &   &   \\
                  F_{1,0} & F_{1,1} & F_{1,9} &   &  &   \\
                    & F_{9,1} & F_{9,9} & F_{9,11} &   &   \\
                    &   &   F_{11,9} & F_{11,11} & F_{11,19} & \\
                    &   &   &   F_{19,11} & F_{19,19} & F_{19,20}\\
                    &   &   &   & F_{20,19} & F_{20,20} \\
                \end{pmatrix}.
\end{equation}
The propagation time from $|0\rangle$ to $|1\rangle$ is
$J_0(1)/J_1(1)$; from $|1\rangle$ to $|9\rangle$ is $J_0(1)/J_8(1)$;
from $|9\rangle$ to $|11\rangle$ is $J_0(1)/J_2(1)$; and so on. The
delocalization time from $|0\rangle$ to $|20\rangle$ is estimated to
be
\begin{equation}
T_{1/q} \approx \frac{J_0(1) J_0(1) J_0(1) J_0(1) J_0(1)}{J_1(1)
J_8(1) J_2(1) J_8(1) J_1(1)} = 1.33\times 10^{15}.
\end{equation}
This is more realistic than Eq. \ref{degenerate1}. Another path of
wave propagation, $|0\rangle \rightarrow |3\rangle \rightarrow
|7\rangle \rightarrow |13\rangle \rightarrow |17\rangle \rightarrow
|20\rangle$, gives
\begin{equation}
T_{1/q} \approx \frac{J_0(1) J_0(1) J_0(1) J_0(1) J_0(1)}{J_3(1)
J_4(1) J_6(1) J_4(1) J_3(1)} = 5.34\times 10^{12},
\end{equation}
which is close to the delocalization time $2^{45} = 3.52\times
10^{13}$. One problem of the degenerate perturbation theory is
$F_{\text{appr}}$ is not unitary.

\emph{An incommensurate case of delocalization.}---The smaller $q$,
the faster the delocalization. If $\tau/2\pi=p/q \approx p'/q'$ and
$q' \ll q$, QKR with $\tau=2\pi p/q$ delocalizes quicker because it
is closer to a stronger resonance. But $\tau=2\pi/q$ is far from any
strong resonance in all the $(\tau= 2\pi p/q)$s, where
$p=1,2,\cdots,q-1, q$. So it has the largest delocalization time and
$T_{1/q} \approx \exp(cq/k)$ is the upper limit of delocalization
time in all the $(\tau=2\pi p/q)$s.

Now we construct irrational $\tau/2\pi$ with delocalization. Imagine
two QKRs with almost equal kick period $\tau$ and $\tau'$ and the
equal kick strength $k$. $\delta\tau=|\tau-\tau'| \ll 1$. $U(M,
\tau)$ is the $M$-period unitary operator with the kick periods
$\tau$ and $U(M, \tau')$ with $\tau'$. The difference between the
matrix elements of two unitary operators \cite{Casati1984,
TaoMa2007GeneralI}
\begin{equation} \label{Lyaponov}
|U(M,\tau)_{nm}-U(M,\tau')_{nm}| \leq \gamma M^{3}k^{2} \delta \tau.
\end{equation}
As the particular value of $\gamma$ is not important, we set
$\gamma=1$. Before $(\epsilon/(k^{2} \delta \tau))^{1/3}$-time
kicks, $|U(M,\tau)_{nm}-U(M,\tau')_{nm}| \leq \epsilon$.

We consider $k=1$ and construct
\begin{equation}\label{irrational1}
\begin{split}
\tau/2\pi =
&1/q+1/\lfloor \exp(3c_1q)\rfloor+1/\lfloor \exp(3c_2\exp(3c_1q))\rfloor\\
&+1/\lfloor \exp(3c_3\exp(3c_2\exp(3c_1q))) \rfloor \\
&+\cdots \\
&+1/\lfloor\exp(3c_n\cdots\exp(3c_3\exp(3c_2\exp(3c_1q)))) \rfloor \\
&+\cdots. \\
\end{split}
\end{equation}
$ \lfloor x \rfloor$ is an integer around the real number $x$ (For
the convenience of the argument below, $ \lfloor x \rfloor$ is not
the same as the floor function in mathematics.) and ensures every
term is a rational number. $q$ is a positive integer such as $20$.
$c_1>c_{1\text{d}}$ and $c_{1\text{d}}$ is the factor in the
delocalization time $T_{1/q}=\exp(c_{1\text{d}}q)$.
$c_2>c_{2\text{d}}$ and $c_{2\text{d}}$ is the factor in the
delocalization time $T_{1/q_1}=\exp(c_{2\text{d}}q_1)$, where
$q_1=\lfloor \exp(3c_1q) \rfloor$. And so on.
\begin{equation}\label{exp3q}
|\tau/2\pi-1/q| \approx 1/ \exp(3c_1q).
\end{equation}
After $\sqrt[3]{q_1}$-time kicks, the dynamics of $\tau$ and
$\tau'=2\pi/q$ will not diverge from each other much due to Eqs.
$(\ref{Lyaponov})$ and $(\ref{exp3q})$. QKR with $\tau$ propagates
to a domain in the momentum space as large as $l_1$. We choose $c_1
\gg {c_{1\text{d}}}$ to ensure $l_1 \gg k^2/4=1/4$. We choose
$\lfloor \exp(3c_1q)\rfloor$ to be an integer approximately
$\exp(3c_1q)$ and to be multiples of $q$. So from $\sqrt[3]{q_1}$
kicks to $\sqrt[3]{q_2}$ kicks, the delocalization speed of QKR is
larger than or equal to QKR with $2\pi/q_1$. After
$\sqrt[3]{q_2}$-time kicks, QKR propagates to a larger domain
$l_2>l_1$. We choose $c_2\gg {c_{2\text{d}}}$ to ensure $l_2 \gg
l_1$. And so on. $l_{\infty}=\infty$. So QKR with the kick period
$\tau$ will delocalize.

Even if $T_{1/q}\neq \exp(cq/k)$, we can always construct
\begin{equation}\label{irrational2}
\frac{\tau}{2\pi} =\frac{1}{q}+\frac{1}{\lfloor T_{1/q}^3
\rfloor}+\frac{1}{\lfloor T_{1/(T_{1/q}^3)}^3\rfloor }+\cdots.
\end{equation}
QKR with $\tau$ in Eq. (\ref{irrational2}) delocalizes.

We note similar irrational numbers have been constructed by Avron
\emph{et al} concerning the Harper equation \cite{avron1982scs} and
by Berry \cite{Berry1984} and Prange \emph{et al} \cite{Prange1982}
concerning the Maryland model. It cannot be a coincidence that
similar numbers are constructed to three totally different problems.
We think such irrational numbers \emph{universally} have similar
behavior with rational numbers in problems of quantum chaos. The way
to construct irrational numbers in Eqs. (\ref{irrational1}) and
(\ref{irrational2}) is very general and our argument depends on Eq.
\ref{Lyaponov}, which is a \emph{universal} quantum Lyaponov
equation \cite{TaoMa2007GeneralI}.


\emph{Problems.}---Some problems remain. First, how does $k$
influence $T_{1/q}$? Second, how to estimate $T_{p/q}$? Do
$T_{p/q}$s generally approximate to $T_{1/q}$? Third, is there one
incommensurate case of delocalization, which is not similar to Eqs.
(\ref{irrational1}) and (\ref{irrational2})? We think localization
happens to the general Liouville number $\tau/2\pi$, such as the
Liouville constant.

\emph{Conclusion.}---First, we have calculated the long time
behavior of QKR using IUMM. It is discovered the delocalization time
is exponentially large for large denominators. Second, we have
constructed an irrational number of delocalization. Concerning QKR,
Eqs. (\ref{irrational1}) and (\ref{irrational2}) are the first
irrational number with delocalization ever known. Both results have
important meaning for the theory of QKR. Third, the large
delocalization time is explained by the degenerate perturbation
theory, which is suggested by and consistent with the delocalization
path of the numerical calculation. The phenomena that the wave
propagates between degenerate or almost degenerate states may be
found in many other systems.

This work is supported by the National Natural Science Foundation of
China under Grant Numbers 10674125 and 10475070. I would like to
thank Professor Fishman for helpful discussions.

\bibliography{IUMM}

\begin{thebibliography}{12}
\expandafter\ifx\csname natexlab\endcsname\relax\def\natexlab#1{#1}\fi
\expandafter\ifx\csname bibnamefont\endcsname\relax
  \def\bibnamefont#1{#1}\fi
\expandafter\ifx\csname bibfnamefont\endcsname\relax
  \def\bibfnamefont#1{#1}\fi
\expandafter\ifx\csname citenamefont\endcsname\relax
  \def\citenamefont#1{#1}\fi
\expandafter\ifx\csname url\endcsname\relax
  \def\url#1{\texttt{#1}}\fi
\expandafter\ifx\csname urlprefix\endcsname\relax\def\urlprefix{URL }\fi
\providecommand{\bibinfo}[2]{#2}
\providecommand{\eprint}[2][]{\url{#2}}

\bibitem[{\citenamefont{{Casati} et~al.}(1979)\citenamefont{{Casati},
  {Chirikov}, {Izraelev}, and {Ford}}}]{Casati1979}
\bibinfo{author}{\bibfnamefont{G.}~\bibnamefont{{Casati}}},
  \bibinfo{author}{\bibfnamefont{B.~V.} \bibnamefont{{Chirikov}}},
  \bibinfo{author}{\bibfnamefont{F.~M.} \bibnamefont{{Izraelev}}},
  \bibnamefont{and} \bibinfo{author}{\bibfnamefont{J.}~\bibnamefont{{Ford}}},
  in \emph{\bibinfo{booktitle}{Stochastic Behavior in Classical and Quantum
  Hamiltonian Systems}}, edited by
  \bibinfo{editor}{\bibfnamefont{G.}~\bibnamefont{{Casati}}} \bibnamefont{and}
  \bibinfo{editor}{\bibfnamefont{J.}~\bibnamefont{{Ford}}}
  (\bibinfo{publisher}{Springer, Berlin}, \bibinfo{year}{1979}),
  vol.~\bibinfo{volume}{93} of \emph{\bibinfo{series}{Lecture Notes in
  Physics}}, pp. \bibinfo{pages}{334--352}.

\bibitem[{\citenamefont{St{\"o}ckmann}(1999)}]{stockmann1999qci}
\bibinfo{author}{\bibfnamefont{H.~J.} \bibnamefont{St{\"o}ckmann}},
  \emph{\bibinfo{title}{{Quantum Chaos: An Introduction}}}
  (\bibinfo{publisher}{Cambridge University Press, Cambridge, England},
  \bibinfo{year}{1999}).

\bibitem[{\citenamefont{Chirikov}(1979)}]{chirikov1979uim}
\bibinfo{author}{\bibfnamefont{B.~V.} \bibnamefont{Chirikov}},
  \bibinfo{journal}{Phys. Rep.} \textbf{\bibinfo{volume}{52}},
  \bibinfo{pages}{263} (\bibinfo{year}{1979}).

\bibitem[{\citenamefont{Greene}(1979)}]{greene1979mds}
\bibinfo{author}{\bibfnamefont{J.~M.} \bibnamefont{Greene}},
  \bibinfo{journal}{J. Math. Phys.} \textbf{\bibinfo{volume}{20}},
  \bibinfo{pages}{1183} (\bibinfo{year}{1979}).

\bibitem[{\citenamefont{Fishman et~al.}(1982)\citenamefont{Fishman, Grempel,
  and Prange}}]{Fishman82}
\bibinfo{author}{\bibfnamefont{S.}~\bibnamefont{Fishman}},
  \bibinfo{author}{\bibfnamefont{D.~R.} \bibnamefont{Grempel}},
  \bibnamefont{and} \bibinfo{author}{\bibfnamefont{R.~E.}
  \bibnamefont{Prange}}, \bibinfo{journal}{Phys. Rev. Lett.}
  \textbf{\bibinfo{volume}{49}}, \bibinfo{pages}{509} (\bibinfo{year}{1982}).

\bibitem[{\citenamefont{Hofstadter}(1976)}]{Hofstadter76}
\bibinfo{author}{\bibfnamefont{D.~R.} \bibnamefont{Hofstadter}},
  \bibinfo{journal}{Phys. Rev. B} \textbf{\bibinfo{volume}{14}},
  \bibinfo{pages}{2239} (\bibinfo{year}{1976}).

\bibitem[{\citenamefont{Casati and Guarneri}(1984)}]{Casati1984}
\bibinfo{author}{\bibfnamefont{G.}~\bibnamefont{Casati}} \bibnamefont{and}
  \bibinfo{author}{\bibfnamefont{I.}~\bibnamefont{Guarneri}},
  \bibinfo{journal}{Commun. Math. Phys.} \textbf{\bibinfo{volume}{95}},
  \bibinfo{pages}{121} (\bibinfo{year}{1984}).

\bibitem[{\citenamefont{Ma}()}]{TaoMa2007GeneralI}
\bibinfo{author}{\bibfnamefont{T.}~\bibnamefont{Ma}}, \eprint{General theory of
  the quantum kicked rotator. I, nlin/0710.1661}.

\bibitem[{\citenamefont{Casati et~al.}(1986)\citenamefont{Casati, Ford,
  Guarneri, and Vivaldi}}]{Casati1986}
\bibinfo{author}{\bibfnamefont{G.}~\bibnamefont{Casati}},
  \bibinfo{author}{\bibfnamefont{J.}~\bibnamefont{Ford}},
  \bibinfo{author}{\bibfnamefont{I.}~\bibnamefont{Guarneri}}, \bibnamefont{and}
  \bibinfo{author}{\bibfnamefont{F.}~\bibnamefont{Vivaldi}},
  \bibinfo{journal}{Phys. Rev. A} \textbf{\bibinfo{volume}{34}},
  \bibinfo{pages}{1413} (\bibinfo{year}{1986}).

\bibitem[{\citenamefont{Avron and Simon}(1982)}]{avron1982scs}
\bibinfo{author}{\bibfnamefont{J.}~\bibnamefont{Avron}} \bibnamefont{and}
  \bibinfo{author}{\bibfnamefont{B.}~\bibnamefont{Simon}},
  \bibinfo{journal}{Bull. Am. Math. Soc.} \textbf{\bibinfo{volume}{6}},
  \bibinfo{pages}{81} (\bibinfo{year}{1982}).

\bibitem[{\citenamefont{Berry}(1984)}]{Berry1984}
\bibinfo{author}{\bibfnamefont{M.~V.} \bibnamefont{Berry}},
  \bibinfo{journal}{Physica} \textbf{\bibinfo{volume}{10D}},
  \bibinfo{pages}{369} (\bibinfo{year}{1984}).

\bibitem[{\citenamefont{Prange et~al.}(1984)\citenamefont{Prange, Grempel, and
  Fishman}}]{Prange1982}
\bibinfo{author}{\bibfnamefont{R.~E.} \bibnamefont{Prange}},
  \bibinfo{author}{\bibfnamefont{D.~R.} \bibnamefont{Grempel}},
  \bibnamefont{and} \bibinfo{author}{\bibfnamefont{S.}~\bibnamefont{Fishman}},
  \bibinfo{journal}{Phys. Rev. B} \textbf{\bibinfo{volume}{29}},
  \bibinfo{pages}{6500} (\bibinfo{year}{1984}).

\end{thebibliography}
\end{document}